\documentclass[twocolumn,showpacs,showkeys,pra]{revtex4-2}
\usepackage{amsmath}
\usepackage{graphicx}
\usepackage{color}
\usepackage{amssymb}
\usepackage{float}

\begin{document}
\title{	
Bloch oscillations of  helicoidal spin-orbit coupled Bose-Einstein condensates  in deep optical lattices}
\author{Sumaita Sultana}
\email{reach2sumaita@gmail.com}
\affiliation{Department of Physics, Kazi Nazrul University, Asansol-713340, W.B., India}
\author{Golam Ali Sekh}
\email{skgolamali@gmail.com}
\affiliation{Department of Physics, Kazi Nazrul University, Asansol-713340, W.B., India}
\begin{abstract}
We consider helicoidal spin-orbit coupled Bose-Einstein condensates in deep optical lattice and study the dynamics of Bloch oscillation.  We show that the variation of helicoidal gauge potential with spin-orbit coupling is different in zero-momentum and plane-wave phases. The characteristics of Bloch oscillation are different in the two phases. In the zero-momentum phase,  the Bloch oscillation is harmonic while it is anharmonic in the plane-wave phase. The amplitude  of Bloch oscillation is found to be affected by the helicoidal gauge potential and spin-orbit coupling. We examine that the  decay of Bloch oscillation caused by mean-field interaction can be managed by helicoidal spin-orbit coupling. 
\end{abstract}

\maketitle
\section{Introduction}
Bloch wave is a delocalized state in lattice potential and it plays a central role  in transport.  This transport decreases drastically in presence of an additional static field $F$ as the force causes localization of the Bloch wave.  This results oscillatory motion of localized Bloch waves, called Bloch oscillation.  For stronger field, Bloch oscillation decays due to coupling with higher bands \cite{r1,r2,r3,r4,r5}. 

Ultra-cold atom in optical lattice is an interesting system for the observation of Bloch oscillation with the difficulties lesser than those faced in solid state physics \cite{r6}. In this system, the effects of atomic interaction come into play through scattering. A breakdown of Bloch oscillation of Bose-Einstein condensates (BECs) occurs due to the nonlinear atomic interaction \cite{r7,r7a,r4}.  In the regime of relatively weak nonlinear interaction, nonlinear dephasing can be considered as a mechanism for the breakdown of Bloch oscillation \cite{r7b}. 

The motion of electron in solid material links with electron’s spin through the spin-orbit coupling (SOC) and thus can influence the Bloch oscillation. However, the SOC is weak and difficult to increase in solid.  In BEC, spin-orbit coupling can be induced and its strength can be increased by Raman lasers. In shallow optical lattice, the effects of SOC have been studied and shown that the two spin states can exhibit out-of-phase oscillation for non-zero Zeeman splitting \cite{r8} and Bloch oscillation is strongly correlated with spin Hall effect \cite{r9}.  By adjusting SOC strength, one can control the direction of motion and can completely suppress the oscillation at a particular value of SOC \cite{r10}. In deep optical lattice, the Bloch oscillation in SOC-BEC can be controlled by Raman coupling. Particularly, it is harmonic in zero-momentum phase while it is anharmonic in plan-wave phase and one can achieve them through proper choice of Raman coupling strength \cite{r11}. 

The strength of SO coupling is tunable with the help of various techniques \cite{r12,r13}, particularly, by creating appropriate artificial gauge potential.  For a helicoidal shaped gauge potential, the SO coupling becomes non-uniform and it is highly tunable \cite{r13,r14,r15,r16}.  The helicoidal SOC  helps propagate soliton over a long distance without attenuation \cite{r16}.  Specifically, it supports the propagation of bright solitons \cite{r16}, stripe solitons \cite{r17}, multi-hump solitons \cite{r18}, Peregrine solitons \cite{r19, r20}, and kink solitons \cite{r21}.  In the recent past, spin-mixing dynamics \cite{r22} and modulational instability \cite{r23} have been studied in helicoidal SOC-BECs.  Recently, the effects of helicoidal gauge potential on Josephson -type oscillation and quantum mechanical self-trapping in presence of optical lattices have also been investigated in the same system \cite{r24}.  However, the effects of spin-orbit coupling on the Bloch oscillation in helicoidal SOC-BEC are not widely explored. 

Our objective in this paper is to study the dynamics of Bloch oscillation in helicoidal spin-orbit coupled Bose-Einstein condensates in deep optical lattice. We deal the system under tight-binding approximation. Within the frame-work of variational method and numerical simulation we study the dynamics of the center of mass in optical lattice \cite{r11}. The center-of-maas executes oscillatory motion if the lattice is tilted (non-zero static force) representing Bloch oscillation (BO). Amplitude of BO diminishes and approaches to zero asymptotically with time. However, a proper management of interaction and spin-orbit coupling parameters can lead to long-lived BO. We identify two typical phases, namely, plane-wave (PW) and zero-momentum (ZM) using fixed point analysis.  In these two phases, variation of helicoidal gauge potential (HGP) with spin-orbit coupling (SOC) are quite different. For  smaller values of SOC, the HGP decreases in ZM phase while it increases in PW phase. The HGP, however, increases in both the phases for larger values of SOC. Like the case of uniform SOC-BEC (without helicoidal gauge potential), the Bloch oscillation is harmonic in zero-momentum phase while it is anharmonic in plane-wave phase \cite{r11}.

In Sec. II, we introduce the theoretical model to describe helicoidal SOC-BEC and then present its discretized version considering tight binding approximation \cite{r11,r25,r26,r27}. Based on variational approach we find equations for the parameters of the condensates wavefunctions.  In section III, we make use of the method of fixed-point analysis and find the variation of spin-orbit coupling and helicoidal gauge potential for both zero momentum and plane wave phases.  In section IV, we discuss dynamics of center of mass for tilted lattice. We also write an effective equation for the center-of-mass to explain cause of decay and possibility  of restoration of Bloch oscillation.  Finally, a brief summary of the work is given in section V.

\section{General Theoretical Formalism}
We consider a Bose-Einstein condensate of $^{87}\mathrm{Rb}$ atoms with total angular momentum $f = 1$.  The SO coupling between  the hyperfine states $\vert f=1, m_f=0\rangle$ and $\vert f=1, m_f=-1\rangle$ are achieved while the $|f=1, m_f=1\rangle$ state is  kept far-detuned from them using  a pair of Raman lasers \cite{r28}. Note that atoms are neutral and their spin-orbit coupling does not come intrinsically in Bose-Einstein condensates. Particularly, the SO coupling is engineered through the generation of artificial gauge potential.  Strength and shape of the SO coupling  can be varied by appropriately taking the gauge potential \cite{r12}. Using a helicoidal-shaped gauge potential one can create non-uniform SOC, called helicoidal spin-orbit coupling in  BECs. Treating  coupled hyperfine states as two pseudo spin-1/2 states and labeling  them by $|\psi_{\uparrow}\rangle$ and $|\psi_{\downarrow}\rangle$, and , taking $m=\hbar=1$,  the helicoidal spin-orbit coupled BECs  can be described by the following Gross-Pitaevskii equation \cite{r17}.  
\begin{eqnarray}
i \psi_t = -\frac{1}{2} \psi_{xx} - i (\alpha \sigma_1 + \kappa \sigma_3) \psi_x \!+ \!\frac{\Delta}{2} \sigma_3 \psi +\Gamma \psi.
\label{eq1}
\end{eqnarray}
Here $\psi=(\psi_{\uparrow}, \psi_{\downarrow})^{T}$ stands for spinor order parameter, $\kappa$ and $\alpha$ are the strengths of helicoidal gauge potential and spin-orbit coupling,   and $\Gamma={\rm diag}(\gamma_1 |\psi_{\uparrow}|^2+\gamma_2  |\psi_{\downarrow}|^2,\gamma_2 |\psi_{\uparrow}|^2+\gamma_1 |\psi_{\downarrow}|^2)$. Here $\gamma_1$ and $\gamma_2$ give intra and inter-atomic interactions. The symbol $\sigma_i$ represents Pauli spin matrix and $\Delta$ gives the Zeeman splitting.
In presence  of titled optical lattice,
\begin{equation}
V(x)=\tilde{F} x+V_0 \cos(2 k x),
\label{eq2}
\end{equation}
 with wave number $k$ and external force $\tilde{F}$, Eq.(\ref{eq1}) can be rewritten as
\begin{eqnarray}
i \psi_t &=& -\frac{1}{2} \psi_{xx} - i (\chi \sigma_1 + \beta \sigma_3) \psi_x + \frac{\Delta  \sigma_3 }{2}\psi \\ \nonumber &+&\Gamma \,\psi+V(x)\psi.
\label{eq3}
\end{eqnarray}

A spin-orbit coupled Bose-Einstein condensate loaded in deep optical lattice can be treated within the framework of tight-binding approximation. {Here we substitute the condensates' wave functions $\psi_{\uparrow}(x,t)$ by $\sum_{n,m} \psi_{\uparrow n}(t)w_m(x-n)$ and $\psi_{\downarrow}(x,t)$ by $\sum_{n,m} \psi_{\downarrow n}(t)w_m(x-n)$ in Eq.(3) and take projection along $w(x-n)$.} This gives the following discrete nonlinear Schr\"odinger equation \cite{r11,r25,r26,r27}.
\begin{eqnarray}
i\frac{\partial \psi_{\uparrow n}}{\partial t}
&=&-J(\psi_{\uparrow n+1}+\psi_{\uparrow n-1})+i\beta (\psi_{\uparrow n+1}-\psi_{\uparrow n-1})\nonumber\\ &-&i\chi (\psi_{\downarrow n+1}-\psi_{\downarrow n-1})+(F\,n+\frac{\Delta}{2})\psi_{\uparrow n} \nonumber\\ &+& (g_1|\psi_{\uparrow n}|^2+g_{2}|\psi_{\downarrow n}|^2)\psi_{\uparrow n},
\label{eq4}
\end{eqnarray}
and
\begin{eqnarray}
i\frac{\partial \psi_{\downarrow n}}{\partial t}
&=&-J(\psi_{\downarrow n+1}+\psi_{\downarrow n-1})-i\beta (\psi_{\downarrow n+1}-\psi_{\downarrow n-1})\nonumber\\ &-&i\chi (\psi_{\uparrow n+1}-\psi_{\uparrow n-1})+(F\,n- \frac{\Delta}{2})\psi_{\downarrow n} \nonumber\\ &+&(g_1|\psi_{\downarrow n}|^2+g_{2}|\psi_{\uparrow n}|^2)\psi_{\downarrow n},
\label{eq5}
\end{eqnarray}
where,
\begin{eqnarray}
J&=&J_{n,n+1}=\int{w^*(x-n)\frac{\partial^2}{\partial x^2}w(x-n-1)dx},\nonumber \\
\chi&=&\chi_{n,n+1}=\frac{2\alpha}{k}\int{w^*(x-n)\frac{\partial}{\partial x}w(x-n-1)dx},\nonumber\\
\beta&=&\beta_{n,n+1}=\frac{2\kappa}{k}\int{w^*(x-n)\frac{\partial}{\partial x}w(x-n-1)dx},\nonumber\\
F&=&\tilde{F}_{n,n+1}=\int{w^*(x-n) xw(x-n-1)dx},\nonumber\\
g_1&=&\!\gamma_1\!\!\int\!|w(x-n)|^4dx, g_2=\!\gamma_2\!\!\int\!|w(x-n)|^4dx.
\label{eq6}
\end{eqnarray}
Here $w(x-n)$ stands for Wannier wave function and it is orthonormal at the lattice site $n$.  {In writing Eqs. (4) and (5), we assume that $J(n,n+1)=J(n,n-1)$, $\chi(n,n)=0$, $\chi(n,n-1)=-\chi(n,n+1)=-\chi(n-1,n)$, $\beta(n,n)=0$, $\beta(n,n-1)=-\beta(n,n+1)=-\beta(n-1,n)$ and restrict the summation over $n$ to on-site and to next neighbor only \cite{ms1,r27}.} In Eq. (\ref{eq6}), $J$, $\beta$ and $\chi$ give dimensionless tunneling coefficient, helicoidal gauge potential and spin-orbit coupling strength respectively. Hamiltonian  density of the system is given by
\begin{eqnarray}
{\cal H}&=&\sum_n \bigg[-J(\psi_{\uparrow n}^*\psi_{\uparrow n+1}+\psi_{\downarrow n}^*\psi_{\downarrow n+1})\nonumber\\ &-&i\chi\psi_{\uparrow n}^*(\psi_{\downarrow n+1}-\psi_{\downarrow n-1})\nonumber\\ &+&i\beta(\psi_{\uparrow n}^*\psi_{\uparrow n+1}-\psi_{\downarrow n}^*\psi_{\downarrow n+1})+c.c\bigg]\nonumber\\&+&\sum_n\bigg[(nF-\frac{\Delta}{2})|\psi_{\downarrow n}|^2+(nF+\frac{\Delta}{2})|\psi_{\uparrow n}|^2\nonumber\\ &+&\frac{1}{2}g_1(|\psi_{\uparrow n}|^4+|\psi_{\downarrow n}|^4)+g_2|\psi_{\uparrow n}|^2|\psi_{\downarrow n}|^2\bigg].
\label{eq7}
\end{eqnarray}
Here $c.c.$ represents complex conjugate of the expression. The Lagrangian density corresponding to $H$ is given by
\begin{eqnarray}
\mathcal{L}=\sum_n \bigg(\frac{i}{2}(\dot{\psi}_{\uparrow n}\psi_{\uparrow n}^*+\dot{\psi}_{\downarrow n}\psi_{\downarrow n}^*)+ c.c.\bigg)-{\cal H}.
\label{eq8}
\end{eqnarray}

Density of a localized state at center is approximately Bell-shaped and, therefore, we chose the following Gaussian trial solution \cite{r29a}
\begin{eqnarray}
\psi_j=\frac{j\sqrt{1+j s}}{\sqrt{2R{\pi}^{1/2}}}e^{-\frac{(n-\xi)^2}{2R^2}+\frac{i\eta(n-\xi)^2}{2}+ip(n-\xi)+j\frac{i\phi}{2}}.
\label{eq12}
\end{eqnarray}
Here $j=\pm 1$ and $\psi_{\uparrow} (\psi_{\downarrow})\equiv \psi_{+1}(\psi_{-1})$. The variational parameters $s$, $R$, $\xi$, $\eta$, $p$ and $\phi$ represent  respectively spin polarization, width, center-of- mass, chirp, momentum and phase. Substituting Eq.(\ref{eq12}) in Eq. (\ref{eq8}) and integrating the resulting equation form $-\infty$ to $+\infty$ with respect to $n$, we get the following Lagrangian.
\begin{eqnarray}
L&=&p\dot{\xi}-\frac{s\dot{\phi}}{2}-\frac{\dot{\eta}R^2}{4}+2Je^{-\sigma}\cos{p}+2\beta s e^{-\sigma}\sin{p}\nonumber\\ &+&2\chi e^{-\sigma}\sqrt{1-s^2}\cos{\phi}\sin{p}-F\xi-\frac{s\Delta}{2}\nonumber\\ &-&\frac{(g_1+g_2)+(g_1-g_2)s^2}{4R\sqrt{2\pi}}.
\label{eq14}
\end{eqnarray}
Here $\sigma=\frac{1}{4}\left(R^{-2}+R^2 \eta^2\right)$. From $\frac{\delta L}{\delta \xi}=0$, $\frac{\delta L}{\delta p}=0$, $\frac{\delta L}{\delta s}=0$, $\frac{\delta L}{\delta R}=0$, and  $\frac{\delta L}{\delta \phi}=0$,
we obtain the following coupled equations.
\begin{eqnarray}
\dot{\xi}&=&2e^{-\sigma}\cos p\left[J\tan p-\beta s-\chi\sqrt{1-s^2}\cos \phi\right],\label{eq15}\\
\dot{p}&=&-F,\label{eq16}\\
\dot{\eta}&=-&\frac{2\sin p(R^4\eta^2-1)}{ e^{\sigma}R^4}\left[J\cot p+\beta s+\chi\sqrt{1-s^2}\cos \phi\right]\nonumber\\ &+&\frac{(g_1+g_2)+(g_1-g_2)s^2}{2\sqrt{2\pi}R^3},\label{eq17}\\
\dot{R}&=&2R\eta e^{-\sigma}\sin p\left[J\cot p+s\beta+\chi\sqrt{1-s^2}\cos \phi\right],\label{eq18}\\
\dot{\phi}&=&-\frac{s(g_1-g_2)}{\sqrt{2\pi }R}-4e^{-\sigma}\sin p\bigg[\frac{s\chi\cos\phi}{\sqrt{1-s^2}}-\beta\bigg]-\Delta,\label{eq19}\\
\dot{s}&=& 4\chi e^{-\sigma}\sqrt{1-s^2}\sin \phi \sin p.\label{eq20}
\end{eqnarray}
We see that the group velocity $\dot{\xi}$ depends on the strengths of helicoidal gauge potential and spin-orbit coupling. The additional force $F$ changes the initial momentum of the wave packet  and thus accelerates it. Eqs. (\ref{eq17}) and (\ref{eq18}) give the dynamics of frequency chirp and width of the wave packet respectively.  Internal Josephson-type oscillation of population imbalance can be described by Eqs. (\ref{eq19}) and (\ref{eq20}).  In following we concentrate on the  dynamics of center-of-mass for $F\neq 0$.

\section{Analysis of plane-wave and zero-momentum phases}
Generally, the energy functional of a spin-orbit coupled BEC contains two degenerate minima for non-zero values of  quasi-momenta and, the condensate wave function is a superposition of the states corresponding to these momentum components. For proper choices of Raman coupling parameter,  the condensate wave function can be reduced to single quasi-momentum component with either zero or non-zero quasi-momentum.  The phase of a SOC-BEC with zero single quasi-momentum  component is called zero-momentum phase while the phase of a SOC-BEC with single but non-zero quasi-momentum component is termed plane-wave phase\cite{r29}.

In order of this, we consider that there is no external force $F=0$ and obtain the ground state solution through the so-called fixed point analysis of the Eqs.(\ref{eq15}) - (\ref{eq20}). Let $\xi_0$, $p_0$, $R_0$, $\eta_0$, $\phi_0$ and $s_0$ are the fixed points for the variational parameters introduced in the trial solution (Eqs.(\ref{eq12})). We make this analysis by considering the fact that the width of the wavepacket at the fixed point ($R=R_0$) is sufficiently large ($R_0>>1$) and its variation ($\dot{R}$) is sufficiently small such that $\eta_0\rightarrow 0$. From Eq.(\ref{eq15}) we get for $\dot{\xi}=0$
\begin{eqnarray}
p_0=\tan^{-1}\left[\frac{\beta s_0+\chi\sqrt{1-s_0^2}\cos{\phi_0}}{J}\right].
\label{eq17a}
\end{eqnarray}
Similarly, from Eqs. (13) and (14)  for $\dot{R}=0$ and $\dot{\eta}=0$ we get,
\begin{eqnarray}
s_0=\sqrt{\frac{g_1+g_2}{g_2-g_1}}.
\label{eq18a}
\end{eqnarray}
For $\dot{s}=0$, we get from Eq.(\ref{eq20})
\begin{eqnarray}
\sin p_0\sin \phi_0=0.
\label{eq19a}
\end{eqnarray}
This implies that  $p_0\neq 0$ and $\phi_0=n\pi$ $(n=0,1,2,\cdots)$. For $\dot{\phi}=0$ we get from Eq. (\ref{eq19})
\begin{eqnarray}
\beta=\frac{s_0 e^{\sigma}(g_1-g_2)}{4\sqrt{2\pi}R_0 \sin{p_0}}+\frac{s_0\cos\phi_0 \chi}{\sqrt{1-s_0^2}}+\frac{\Delta e^{\sigma}}{4 \sin{p_0}}.
\label{eq20a}
\end{eqnarray}
In zero-momentum phase $p_0 \rightarrow 0$ and
$\tan p_0\approx p_0\approx \sin p_0$. Therefore, we get from Eq. (20)
\begin{eqnarray}
\beta=-\frac{c_2\chi}{2}+\frac{\sqrt{4c_1 J+(c_2^2+\cos^2\phi_0)\chi^2}}{2},
\label{eq21a}
\end{eqnarray}
where
\begin{eqnarray}
c_1&=&\left(\frac{ e^{\sigma}(g_1-g_2)}{4\sqrt{2\pi}R_0 }+\frac{\Delta e^{\sigma}}{4s_0}\right),\\
c_2&=&\frac{1}{s_0}\sqrt{1-s_0^2}\cos\phi_0-\frac{s_0\cos\phi_0 }{\sqrt{1-s_0^2}}.
\end{eqnarray}

For plane-wave phase, $p_0 > 0$. In this case $\beta$ and $\chi$ satisfy the following relation
\begin{eqnarray}
(\beta+c_3\chi)^2((\beta+c_4\chi)^2-s_0c_1^2)-J^2c_1^2=0
\end{eqnarray}
where,
\begin{eqnarray}
c_3=\frac{1}{s_0}\sqrt{1-s_0^2}\cos\phi_0,\,\,{\rm and}\,\,c_4=\frac{s_0\cos\phi_0}{\sqrt{1-s_0^2}}.
\end{eqnarray}
\begin{figure}[h!]
\includegraphics[scale=0.5]{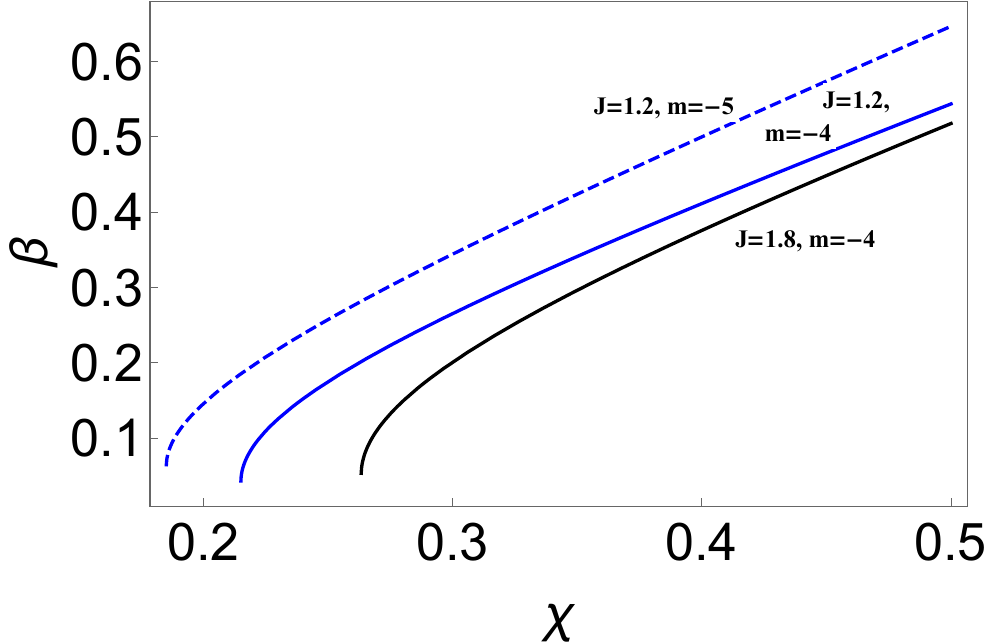}
\caption{Variation of $\beta$ with $\chi$ for different values of $J$ and relative interaction strength $m=g_2/g_1$ in zero-momentum phase. The blue-solid, blue-dashed and black curves are drawn for $\{J,m\}=\{1.2, -4\}$, $\{1.2, -5\}$ and $\{1.8,-4\}$ respectively. Other parameters are taken as $\phi_0=\pi/12$, $R_0=40$, $\Delta=0$ and  $F=0.1$ for all the curves.}
\label{fig1}
\end{figure}
{Understandably, Eqs. (21) and (24) give relations between $\beta$ and $\chi$  in zero-momentum and plane-wave phases respectively.}

In Fig. {\ref{fig1}} we show the variation of the strengths of helicoidal gauge potential ($\beta$) and spin-orbit coupling ($\chi$) in the zero-momentum phase. In this phase, $\beta$ increases as $\chi$ increases. The slope of the curve depends on the tunneling coefficient $J$ and relative interaction parameter $m$, where $m=g_2/g_1$. Particularly, the slope decreases with the increase for both $J$ and $m$.  For chosen values of tunneling coefficient and interaction, therefore, the ratio of helicoidal gauge potential and spin-orbit coupling plays a significant role in studying zero momentum phase.
\begin{figure}[h!]
\includegraphics[scale=0.6]{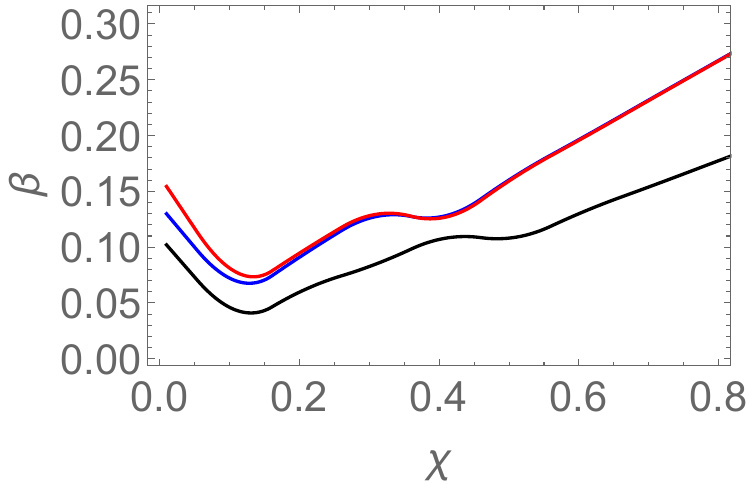}
\caption{Variation of $\beta$ with $\chi$ in plane-wave phase for different values of $J$ and $m(=g_2/g_1)$. Here blue, red and black curves are drawn for $\{J=0.6, m=-1.25$\}, $\{J=0.8, m=-1.25\}$ and $\{J=0.6, m=-1.10\}$ respectively. Here we take $\phi=0$, $R=40$ and $\Delta=0$ and $F=0.1$.}
\label{fig2}
\end{figure}

In the plane-wave phase $\chi$ versus $\beta$ curve shows some interesting features (Fig.2). Particularly, the value of $\beta$ decreases with the increase of $\chi$ and then increases after attending a minimum value. In this phase, the slope of a curve depends  sensitively  on the relative strength $m$ of atomic interactions. Particularly, the slope decreases as the value of $m$ deceases. In contrast to the zero-momentum phase, $\chi-\beta$ curve does not depend sensitively  on the tunneling coefficient $J$ in the plane-wave phase.

{In order to understand the role of helicoidal SOC in determining the phases of the coupled system, we draw again $\chi-\beta$ diagram using Eqs. (21) and (24) keeping other parameters same for the two phases (Fig. 3 ).  It clearly shows that the system prefers to stay in zero-momentum phase if $\beta$ is small while one can observe plane-wave phase for relatively larger values of $\beta$.  Particularly, the ratio of $\beta$ and  $\chi$ is smaller in zero-momentum phase than that in plane-wave phase.}
\begin{figure}[h!]
\includegraphics[scale=0.6]{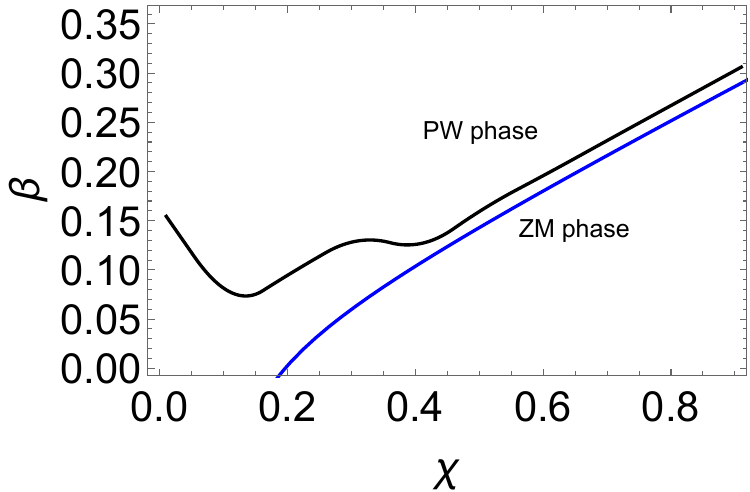}
\caption{Variation of $\beta$ with $\chi$ in plane-wave phase (black curve)  and zero-momentum phase  (blue curve). Here we take $J=0.8$, $g_1=- 8$,  $g_2=10$, $\Delta=0$, $R=40$ and  $F=0.1$.}
\label{fig2}
\end{figure}

\section{Bloch oscillation  in presence of  helicoidal spin-orbit coupling}
From Eq.(13) we see that, for a large value of $R$, $\dot{\eta}\to 0$ and thus $\eta$ can either be zero or constant. For $\eta\to 0$ Eq.(14) gives $\dot{R}\to 0$. Therefore, the variations of $\eta$ and $R$ have negligible effects on the dynamics of $\xi$. In view of this, we first solve Eq. (\ref{eq15}), (\ref{eq16}), (\ref{eq19}) and (\ref{eq20}) numerically for zero-momentum (ZM) phase.  {Particularly, we fix values of different parameters with the help of Eq. (21) and  see that the center-of-mass of the wavepacket executes harmonic motion in presence of external force. This is termed as harmonic Bloch oscillation (BO) (Fig.4). We note that a two-component BEC without helicoidal spin-orbit coupling ($\beta=\chi=0$) in deep optical lattice exhibits Bloch oscillation \cite{r33,r34}. Due to presence of helicoidal SOC, a coupling between the center-of-mass motion and spin degrees of freedom occurs and thus results enhancement of amplitude of BO in the ZM phase}. In order to check the merit of the variational approach, we solve Eqs. (4) and (5) using  fourth-order Runge-Kutta method  by taking time step size $0.0001$ and equilibrium state as initial state.  We see that numerical results accurately describes the harmonic Bloch oscillation in zero-momentum phase and also the results obtained direct numerical simulation  and variational approach shows good agreement.
\begin{figure}[h!]
\begin{center}
\includegraphics[scale=0.6]{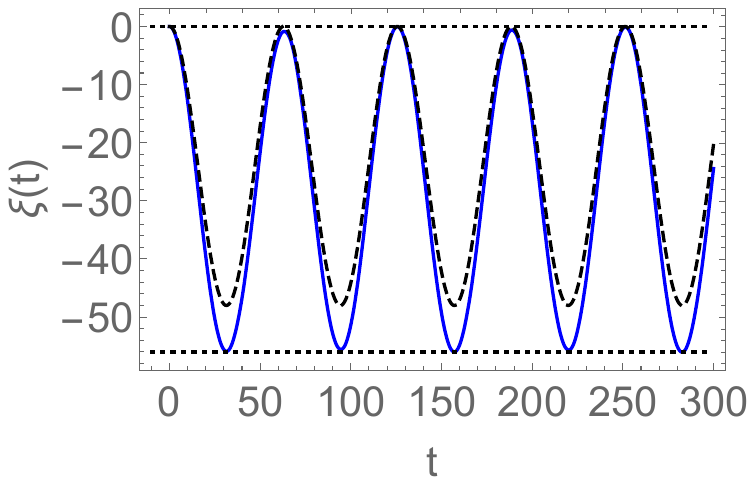}
\includegraphics[scale=0.6]{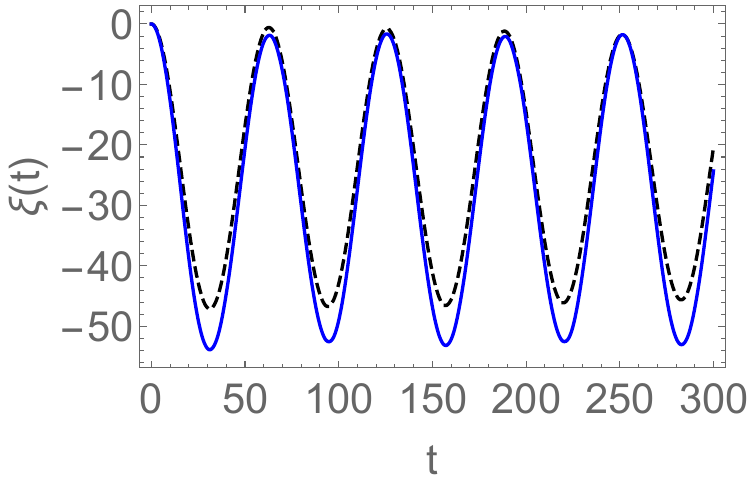}
\caption{Bloch oscillation in zero-momentum phase. 
Top panel: Solid-blue and dashed-black curves are drawn for   $\{\chi = 0.49,   \beta =0.54\}$ and $\{\chi = 0,\beta = 0\}$ respectively. We take other parameters as: $J=1.2$, $m=-4$, $\Delta=0$, $\phi=\pi/12$ and $F=0.1$.
Bottom panel: Oscillation of center-of-mass obtained from direct numerical simulation of coupled discrete nonlinear Schr\"odinger equations in (4) and (5) for  the values of parameters equal to those used for variational results in the top panel.}
\label{fig3}
\end{center}
\end{figure}

\begin{figure}[h!]
\begin{center}
\includegraphics[scale=0.6]{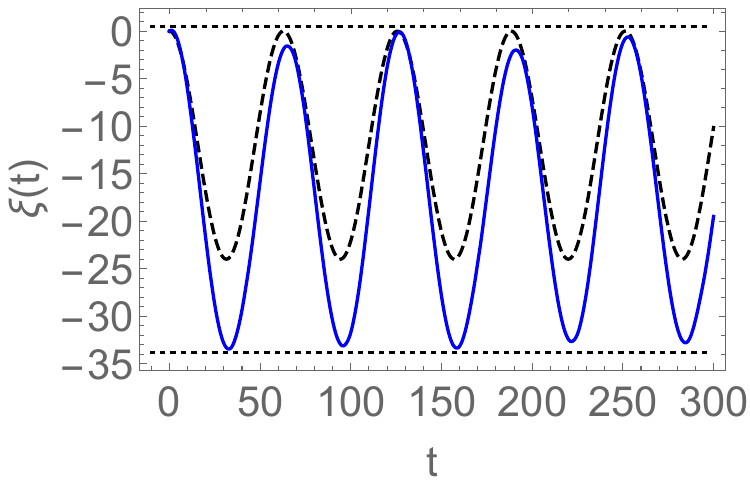}
\includegraphics[scale=0.6]{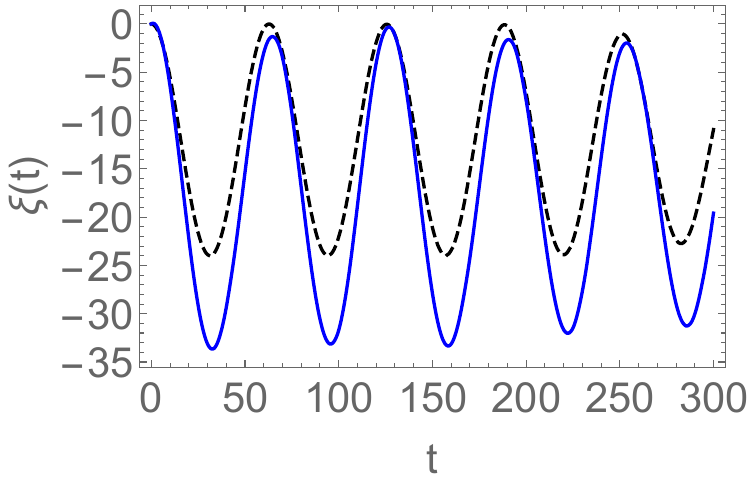}
\caption{It shows Bloch oscillation in plane-wave phase obtained from variational calculation. Top panel: The solid-blue and dashed-black curves are drawn for $\{\chi = 0.57, \beta = 0.20\}$,  and  $\{\chi=0, \beta=0\}$ respectively. We take other parameter as: $J=0.6$, $m=-4$, $\Delta=0$, $\phi=0$ and $F=0.1$. Bottom panel: Oscillation of center-of-mass obtained from direct numerical simulation of coupled discrete nonlinear Schr\"odinger equations in (4) and (5) for  the values of parameters equal to those used for variational results in the top panel.}
\label{fig4}
\end{center}
\end{figure}

{For the plane-wave phase we fix values of different parameters from Eq.(24) and solve Eqs.(11), (12), (15) and (16) numerically.  The variation of center-of-mass with time in presence of external force is shown in Fig. \ref{fig4}. Looking closely into the figure we see that the oscillation of center-of-mass is not exactly harmonic. Particularly, the amplitude of oscillation is changing with time. We term it as anhamonic Bloch oscillation. Comparing with the BO in a binary BEC without helicoidal SOC, we see that the amplitude of BO is augmented also in this phase. }
\begin{figure}[h!]
\includegraphics[scale=0.7]{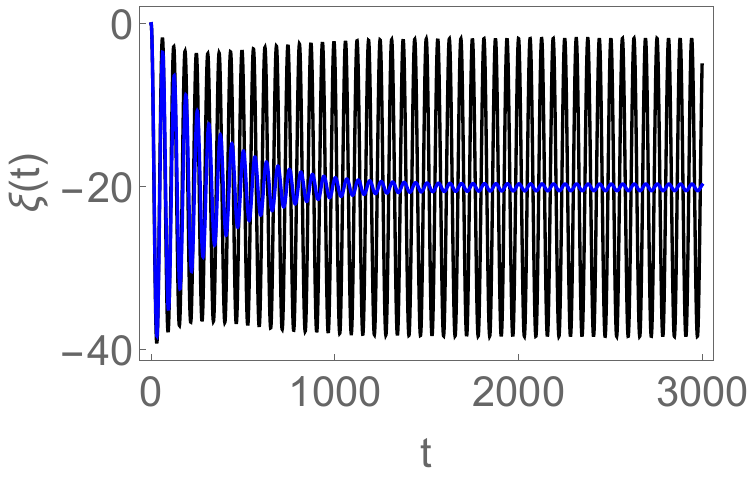}
\caption{Bloch Oscillation damping and restoration. The blue and black curves are drawn for $\chi=0.4$ (decaying BO) and  $\chi=0.9$ (long-lived BO) respectively. Other parameters are taken as: $m=-4$, $J=1.2$, $\beta=0.8$ and $F=0.1$.}
\label{fig5}
\end{figure}

{We have seen that the Bloch oscillation is affected by helicoidal gauge potential due to coupling between the center-of-mass motion of BEC and its spin dynamics \cite{r33a,r33b}. This coupling, particularly, leads to two different phases of the BO. The mean-field atomic interaction can naturally cause damping or even result complete break down of the BO \cite{r7b,r33,r34}.  In order to visualize the effects of helicoidal spin-orbit coupling on the damping of BO, we rewrite the equation of center-of-mass using Eqs. (11) -(16) as follows. } 
\begin{eqnarray}
\frac{d^2\xi}{dt^2}+\gamma\, \frac{d\xi}{dt}+F^2\xi=H_1 F+G \cos(Ft-p_0),
\label{eq26}
\end{eqnarray}
where
\begin{eqnarray}
H_1&=& \left(H-\frac{D}{2}\right),\,\,\gamma={\eta_0 D/2 }\\
H&=&D_1 \sin{p_0}+D_2\cos{p_0}+F\xi_0+\frac{1}{2}(D+s_0\Delta),\\
D&=&\frac{(g_1+g_2)+(g_1-g_2)s_0^2}{2R\sqrt{2\pi}},\,\,D_2=-2 J e^{-\sigma},\\
G\!&=&-\!2\chi\sqrt{1-s_0^2}\,\sin\phi_0\bigg(\!\frac{s_0 (g_1-g_2)}{\sqrt{2\pi}R}\!+\!\Delta\!\bigg),\\
D_1\!&=&\!-2e^{-\sigma}(\chi \sqrt{1-s_0^2}\cos{\phi_0}+\beta s_0).
\end{eqnarray}
{Eq.(\ref{eq26}) represents a standard damped-driven oscillator. Here we assume that the values of population imbalance ($s$), frequency chirp ($\eta$), phase ($\phi$) and width ($R$) do not vary appreciably with time such that the Hamiltonian $H$ remains approximately constant. From Eqs. (26), (27) and (29) we see that the damping of BO depends on the strengths of intra-atomic and inter-atomic interactions of the condensates. Particularly, the amplitude of BO attenuates rapidly for repulsive inter-component interaction \cite{r33,r34}. However, the external force in conjunction with helicoidal spin-orbit coupling produces an effective restoring force to protect the decay of BO.   In Fig. 6 we plot Bloch oscillation for fixed values of inter- and intra-component interactions. It shows that Bloch oscillation diminishes quickly if  spin-orbit coupling is weak (blue curve). However, one can generate relatively long-lived Bloch oscillation by properly tuning the spin-orbit coupling (black curve).} 


\section{conclusions}
Bose-Einstein condensate is a system that can be used to explore different phenomena of condensed matter physics with great flexibility. Bloch oscillation is one of them. Here we consider helicoidal spin-orbit coupled Bose-Einstein condensate in deep optical lattice with a view to study the effects of helicoidal gauge potential and spin-orbit coupling on the Bloch oscillation. We know that zero-momentum phase  and plane-wave phase  are the two typical phases of spin-orbit coupled Bose-Einstein condensate. In the former phase, the two quasi-momentum components of SOC-BEC merge to a single and zero  quasi-momentum state while they merge to a single but non-zero  quasi-momentum state. We have seen that the variations of gauge potential with spin-orbit coupling  in these two phases are different. Particularly,  the variations show opposite trends in the limit of small SOC strength. The BEC is found to execute Bloch oscillation in both the phases. However, their characteristics are different. In plane-wave phase, the Bloch oscillation is anharmonic while it is harmonic in zero-momentum phase.  {We compare the Bloch oscillation of SOC-BEC with that  of two-component BEC without SOC in deep optical lattice and show that helicoidal SOC increases the amplitude of Bloch oscillation in both the phases.}

The amplitude of Bloch oscillation decreases due to presence of mean-field interaction. We have examined that the reduction of amplitude depends on the relative values of inter- and intra-component interaction. For a fixed value of  relative interaction, the Bloch oscillation is again influenced by the helicoidal gauge potential and spin-orbit coupling. Particularly, the amplitude center-of-mass oscillation increases and restores its initial value if the relative value of helicoidal gauge potential and spin-orbit coupling is properly tuned and thus gives a long-lived Bloch oscillation.

\section*{Acknowledgements}
S. Sultana would like to thank  `West Bengal Higher Education Department' for providing Swami Vivekananda Merit-Cum-Means Scholarship with F. No. WBP201637916084.

\end{document}